# Teaching Earth Dynamics: What's Wrong with Plate Tectonics Theory?

J. Marvin Herndon

Transdyne Corporation
San Diego, California 92131 USA

September 30, 2005

## Abstract

Textbooks frequently extol plate tectonics theory without questioning what might be wrong with the theory or without discussing a competitive theory. How can students be taught to challenge popular ideas when they are only presented a one-sided view? In a just a few pages, I describe more than a century of geodynamic ideas. I review what is wrong with plate tectonics theory and with Earth expansion theory, and describe my new *Whole-Earth Decompression Dynamics Theory*, which unifies the two previous, dominant theories in a self-consistent manner. Along the way, I disclose details of what real science is all about, details all too often absent in textbooks and classroom discussions. In these few pages, I only touch on highlights and just part the curtain a bit so that teachers might glimpse ways to bring to their students some of the richness and excitement of discovery that becomes evident when one begins to question prevailing, currently popular perceptions of our world.

## Introduction

Try to answer the question, "What's wrong with plate tectonics theory?" and you begin to get down to the basics of real science. Textbooks often extol the virtues of plate tectonics, especially the apparently obvious agreement between theory and observation, but rarely address the question of what might be wrong with the theory. Posing that question to students may serve as a starting point for discussions that offer the possibility of beginning to address even more fundamental questions in science, like "What is the purpose of science?", "What determines correctness in science?", and "How do new ideas originate?".

Not only do textbooks almost universally fail to address the question of what might be wrong with plate tectonics theory, they rarely, if ever, mention a competitive theory that actually predates plate tectonics, but that has fewer adherents, called Earth expansion theory. How can students be taught to question prevailing, currently popular theories if they are only presented a one-sided view? Questioning and challenging prevailing, popular theories is what science is all about. Remember, popularity only measures popularity, not scientific correctness: Science is a logical process, not a democratic process. If the world really is the way we presently perceive it to be, then there would no need for science. Science is about new ideas, it is about discovery. Science is about understanding more precisely the nature of the Earth and the Cosmos. Asking the question, "What's wrong with plate tectonics theory?" may also help to prepare students to understand and perhaps try to challenge a new idea that supersedes plate tectonics, called *Whole-Earth Decompression Dynamics*.

## A Bit of Historical Background

During the 19th Century, scientists recognized that opposing margins of continents appear to fit together in certain ways and to display geological and ancient-biological evidence of having been joined in the past. Eduard Suess (1831-1914) envisioned a vast super-continent that had broken up in the past with parts of it sinking into the ocean floor. Suess' concept was subsequently recognized to be wrong because continental rock is less dense than ocean floor basalt rock and thus would not be able to sink beneath the ocean floor.

Early in the 20th Century, Alfred Wegener (1880-1930) also proposed that the continents at one time had been united, but subsequently had separated and drifted through the ocean floor to their present positions. Wegener's theory of continental drift was systematically ignored for fifty years. Whoa! What's wrong here? Science is about controversy, about debate. It is about new ideas. Ignoring an important new contradiction to prevailing ideas is really bad science. In fact, it is anti-science. Any important, promising new scientific idea should be discussed and debated. If possible, experiments or theoretical considerations should be made to test the idea. If the new idea is found to be wrong, it should be refuted in the scientific literature, ideally in the journal where it was originally published; otherwise, it should be acknowledged. New ideas often beget other new ideas and science advances. Early on, students should learn about this process and should be exposed to the excitement and debate that is a natural part of good science.

## A Questioning Look at Plate Tectonics Theory

Finally, after half a century of silence, debate erupted in the 1960s. Alfred Wegener's idea of continental drift re-emerged, cast into a new form called plate tectonics theory, with more detail and with new supporting observational data. Plate tectonics is based upon the idea that the crust of the Earth is composed of a set of movable plates that are in motion. The theory also embraces the idea of seafloor spreading.

In plate tectonics theory, basalt rock, observed erupting from the mid-oceanic ridges, is thought to creep slowly across the ocean basin and to subduct, to plunge into the Earth, typically into submarine trenches. Indeed, there is convincing evidence for such ocean floor movement. When basalt rock cools, its magnetic minerals act like magnetic tape recorders, locking in a record of the prevailing direction of the Earth's magnetic field. The observed pattern of magnetic stripes across the ocean floor reflects reversals of the geomagnetic field, recorded as new seafloor cooled and moved away from the mid-oceanic ridge. This might appear to be convincing evidence of plate tectonics theory. But does that mean plate tectonics theory is totally correct? Not necessarily. Look deeper and look questioningly.

Imagine new seafloor being produced at the mid-oceanic ridges, moving away slowly, and ultimately plunging into the Earth (subducting). If the Earth is not getting any larger, then what is happening to all that seafloor that is plunging downward? According to plate tectonics theory, the Earth's mantle is convecting, turning over and over in an endless loop. But is it? Is it really? Is there any direct evidence that the mantle is convecting?

The idea of convection comes from what is observed when liquids are heated from beneath. The heated liquid expands a little, becoming slightly less dense than the cooler liquid and thus tends to rise, to float to the surface. Cool liquid rushes in to fill the space that had been occupied by the heated liquid. In a large volume, a geometric pattern of (typically hexagonal-shaped) convection cells often develops.

Convection is assumed to take place in the mantle as solid-state flow instead of liquid-state flow. But there is no indication at the Earth's surface of the formation of convection cells. Indeed, decades of research has yet to find any direct, unambiguous evidence of mantle convection. Is the idea of mantle convection simply a contrivance to make plate tectonics seem to work? Is there another way of looking at the dynamics of the Earth? These are serious questions to consider.

**A Questioning Look at Earth Expansion Theory**

In 1933, Otto Hilgenberg (1896-1976) published a different idea about the continents. He imagined that initially the Earth was smaller in diameter, without oceans, and that the continents formed a uniform shell of matter covering the entire surface of the planet. Hilgenberg's idea was that the Earth subsequently expanded, fragmenting the uniform shell of matter into the continents, and creating ocean basins in between. He spent the rest of his life making impressive model globes, writing, and lecturing on the subject, thus attracting others, such as S. Warren Carey (1911-2002), who built upon the idea and helped to promote it. But there are some really big problems.

One big problem is energy: For the Earth to have expanded it would have to do work against gravity and that would take a HUGE amount of energy, more than all of the Earth's nuclear fuel could have ever have yielded. Such a powerful energy source was unknown. Some scientists even suggested that the expansion came about as a consequence of changing laws of nature, but there is still no independent corroborating evidence.

Another big problem is time scale: The oldest seafloor is only 200 million years old, but the age of the Earth is about 4½ billion years. Earth expansion theory envisions ocean basins being formed as the Earth expanded and thus separated the continents. That would mean, accordingly, that all Earth expansion occurred during the last 200 million years as no seafloor is any older than that. How can that be reasonably explained?

Still another big problem comes from Earth-orbiting satellite data: Modern measurements show that the Earth is presently not expanding. If, according to Earth expansion theory, all expansion took place during the last 200 million years, it would seem strange indeed that there is no present Earth expansion.

Just as in plate tectonics theory, there are appealing aspects to Earth expansion theory, like the origin of continents from a uniform continental shell, but there are also profound difficulties. The fact that decades have passed without reconciliation of these two seemingly disparate scientific theories might suggest that neither is wholly correct. Perhaps ultimately the correct theory lies somewhere between the two. Perhaps the middle-ground is the new theory that I have proposed, called *Whole-Earth Decompression Dynamics*, which unifies certain elements of plate tectonics with certain elements of Earth expansion.

## A Brief Look at *Whole-Earth Decompression Dynamics Theory*
*Current Science (India)*, in press; also posted at
http://UnderstandEarth.com/WEDD.pdf ) and http://arXiv.org/astro-ph/0507001

Think about the planets of our Solar System. Earth, like the other inner planets, consists of rock with a heavy iron-alloy core. The giant outer planets, on the other hand, presumably have rock-plus-alloy interiors, but these are surrounded by great shells of light gases, mainly hydrogen and helium, which are many times more massive than the underlying rock-plus-alloy.

Scientists have spectrographically analyzed sunlight and found that the light elements, like hydrogen and helium, in the outer part of the Sun, are about 300 times more massive than the corresponding heavy rock-and-alloy-forming elements. There are really good reasons to believe that the inner planets formed originally from matter like that in the outer part of the Sun.

There have long been mainly two ideas about how the planets of the Solar System formed. In the 1940s and 1950s the idea was discussed about planets "raining out" from inside of giant gaseous protoplanets. But that idea fell out of fashion and scientists began thinking of the primordial matter, not being dense protoplanets, but rather spread out into a very low-density "solar nebula".

The idea of low-density planetary formation envisioned that dust would condense at fairly low temperatures, and then gather into progressively larger grains, and become rocks, then planetesimals, and ultimately planets. The gaseous components would just go away in an unspecified manner. This is the prevailing, popular view of planetary formation. But remember, popularity only measures popularity, not scientific correctness. So how can you know whether an idea is right or wrong? One way is to look for a contradiction, a consequence that is in conflict with what is observed.

I made thermodynamic calculations on the nature of matter that would condense under such circumstances at low-temperatures/low-pressures and I discovered a major contradiction: Most of the iron would end up being combined with oxygen. Calculations show that there would be insufficient iron metal to account for the massive cores of the inner planets, that are indicated by their bulk densities. So it's back to square one, time to reconsider the "out of fashion" idea of planets "raining out" from inside of giant gaseous protoplanets.

Imagine reconstituting the Earth with all of its primordial light gases that were originally lost during the time the Solar System formed. You would thus be imagining a planet similar in mass to Jupiter, roughly 300 Earth-masses. What would the Earth be like, surrounded by all that gaseous mass? Calculations show that it would be compressed to about 64% of its present diameter. Its surface area would be quite similar to the surface area presently occupied by the continents. In other words, the Earth would have a uniform shell of continental matter covering its entire surface, just as first envisioned by Otto Hilgenberg.

It is known that, at an early time during the formation of the Solar System, the gaseous components of primordial matter were stripped from the inner planets, not just the hydrogen and helium, but even the heavy gases like xenon. Astronomers have observed that very young stars are often quite unstable, erupting with energetic outbursts, shedding matter into space as a super-intense solar wind. It seems reasonable to presume that such outbursts from our young Sun stripped the gaseous envelopes from the rocky inner planets.

Stripped of its great overburden of primordial gases, the compressed Earth would begin to decompress, like a ball that has been squeezed and then released. So, there in nature is a mechanism and a primary energy source for whole-Earth decompression, the stored energy of protoplanetary compression.

The initial whole-Earth decompression is expected to result in a global system of major primary cracks appearing in the rigid crust which persist and are identified as the global, mid-oceanic ridge system, just as explained by Earth expansion theory. But here the similarity with that theory ends. *Whole-Earth Decompression Dynamics Theory* sets forth a different mechanism for whole Earth dynamics which involves the formation of *secondary* decompression cracks and the in-filling of those cracks.

As the Earth subsequently decompresses and swells from within, the deep interior shells may be expected to adjust to changes in radius and curvature by plastic deformation. As the Earth decompresses, the area of the Earth's rigid surface increases by the formation of *secondary* decompression cracks often located near the continental margins and presently identified as submarine trenches. These secondary decompression cracks are subsequently in-filled with basalt, extruded from the mid-oceanic ridges, which traverses the ocean floor by gravitational creep, ultimately plunging into secondary decompression cracks, thus emulating subduction.

As viewed today from the Earth's surface, the consequences of *Whole-Earth Decompression Dynamics* appear very similar to those of plate tectonics, but with some profound differences. In fact, most of the evidence usually presented in support of plate tectonics also supports *Whole-Earth Decompression Dynamics*. Just as in plate tectonics, one sees seafloor being produced at the mid-oceanic ridge, slowly moving across the ocean basin and disappearing into the Earth. But unlike plate tectonics, the basalt rock is not being re-cycled endlessly by convection; instead, it is simply in-filling secondary decompression cracks. From the surface it may be very difficult indeed to discriminate between plate tectonics and *Whole-Earth Decompression Dynamics*. But what about satellite data?

After 4½ billion years, the Earth appears to be approaching the end of its decompression. Satellite length-of-day measurements show virtually no current lengthening, implying no current secondary decompression crack formation. The formation of secondary decompression cracks might be episodic, though, like the release of stress by major earthquakes, or secondary crack formation may have ended forever. But major decompression cracks are still conspicuously evident, for example, circum-pacific submarine trenches, such as the Mariana Trench. And, the complementary *Whole-Earth Decompression Dynamics* process of basalt extrusion and crack in-filling, however, continues at present, but its rate is quite slow, consistent with reported length-of-day measurements.

## Summary

In a just a few pages, I have covered more than a century of geodynamic ideas, reviewed what is wrong with plate tectonics theory and with Earth expansion theory, and I have described my new *Whole-Earth Decompression Dynamics Theory*, which unifies the two previous, dominant theories in a self-consistent manner. Along the way, I have disclosed details of what real science is all about, details all too often absent in textbooks and absent in oral classroom discussions. In these few pages I have only touched on highlights and just parted the curtain a bit so that teachers might glimpse ways to bring to their students some of the richness and excitement of discovery that becomes evident when one begins to question prevailing, currently popular perceptions of our world.